\documentclass{LT23auth}
\usepackage{graphicx}

\newcommand{\PrOsSb}{PrOs$_4$Sb$_{12}$}

\begin{document}

\hyphenation {Schott-ky} \hyphenation {su-per-con-duc-tor}
\hyphenation {su-per-con-duc-ting} \hyphenation
{su-per-con-duc-tiv-i-ty} \hyphenation {quad-ru-pole} \hyphenation
{quad-ru-po-lar} \hyphenation {skut-ter-u-dite} \hyphenation
{an-ti-fer-ro-mag-net-ic} \hyphenation {pre-fac-tor} \hyphenation
{mul-ti-plet}

\begin{frontmatter}

\title{Heavy fermion superconductivity in the filled skutterudite
compound \PrOsSb}

\author{M. B. Maple\thanksref{thank1}},
\author{P.-C. Ho, V. S. Zapf, W. M. Yuhasz,
N. A. Frederick, E. D. Bauer}

\address{Department of Physics and Institute for Pure and
Applied Physical Sciences, University of California, San Diego,
La Jolla, CA 92093 USA}


\thanks[thank1]{Corresponding author. E-mail:mbmaple@physics.ucsd.edu}

\begin{abstract}

The filled skutterudite compound \PrOsSb{} has been found to
exhibit superconductivity with a critical temperature
$T_\mathrm{c} = 1.85$ K that develops out of a heavy Fermi liquid
with an effective mass $m^{*} \approx 50~m_\mathrm{e}$.  The
current experimental situation regarding the heavy fermion state,
the superconducting state, and a high field, low temperature phase
that is apparently associated with magnetic or quadrupolar order
in \PrOsSb{} is briefly reviewed herein.

\end{abstract}

%
%
\end{frontmatter}

\section{Introduction}

About a year ago, we reported that the compound \PrOsSb{} exhibits
heavy fermion superconductivity below a critical temperature
$T_\mathrm{c} = 1.85$ K \cite{Maple01,Bauer02}.  The heavy fermion
state is characterized by an effective mass $m^{*} \approx
50~m_\mathrm{e}$, where $m_\mathrm{e}$ is the free electron mass.
To our knowledge, this is the first heavy fermion superconductor
based on Pr; all of the other known heavy fermion superconductors
(about $20$) are compounds of Ce or U. In an effort to obtain
information about the interactions that are responsible for the
heavy fermion state and the superconductivity of \PrOsSb, we have
performed measurements of certain normal and superconducting state
properties of this compound as a function of temperature,
pressure, and magnetic field \cite{Maple01,Bauer02,Maple02,Ho01}.
An analysis of magnetic susceptibility $\chi{}(T)$, specific heat
$C(T)$, and inelastic neutron scattering INS measurements on
\PrOsSb{} within the context of a cubic crystalline electric field
(CEF) yielded a Pr$^{3+}$ energy level scheme with a $\Gamma_{3}$
nonmagnetic doublet ground state that carries an electric
quadrupole moment, a low lying $\Gamma_{5}$ triplet excited state
at $\sim 10$ K above the ground state, and $\Gamma_{4}$ triplet
and $\Gamma_{1}$ singlet excited states at much higher energies
($\sim 130$ K and $\sim 313$ K, respectively).  This scenario
suggests that the underlying mechanism of the heavy fermion
behavior in \PrOsSb{} may involve electric quadrupole
fluctuations, rather than magnetic dipole fluctuations.  It also
raises the possibility that electric quadrupole fluctuations play
a role in the superconductivity of \PrOsSb. In magnetic fields
greater than $4.5$ T and at temperatures below $1.5$ K, we have
also found evidence for the existence of a region in which there
is magnetic or quadrupolar order \cite{Maple02,Ho01}.  This
suggests that the superconducting phase may occur in the vicinity
of a magnetic or quadrupolar quantum critical point (QCP).

\section{Evidence for heavy fermion superconductivity in \PrOsSb}

\begin{figure}[t]
\begin{center}\leavevmode
\includegraphics[width=0.8\linewidth]{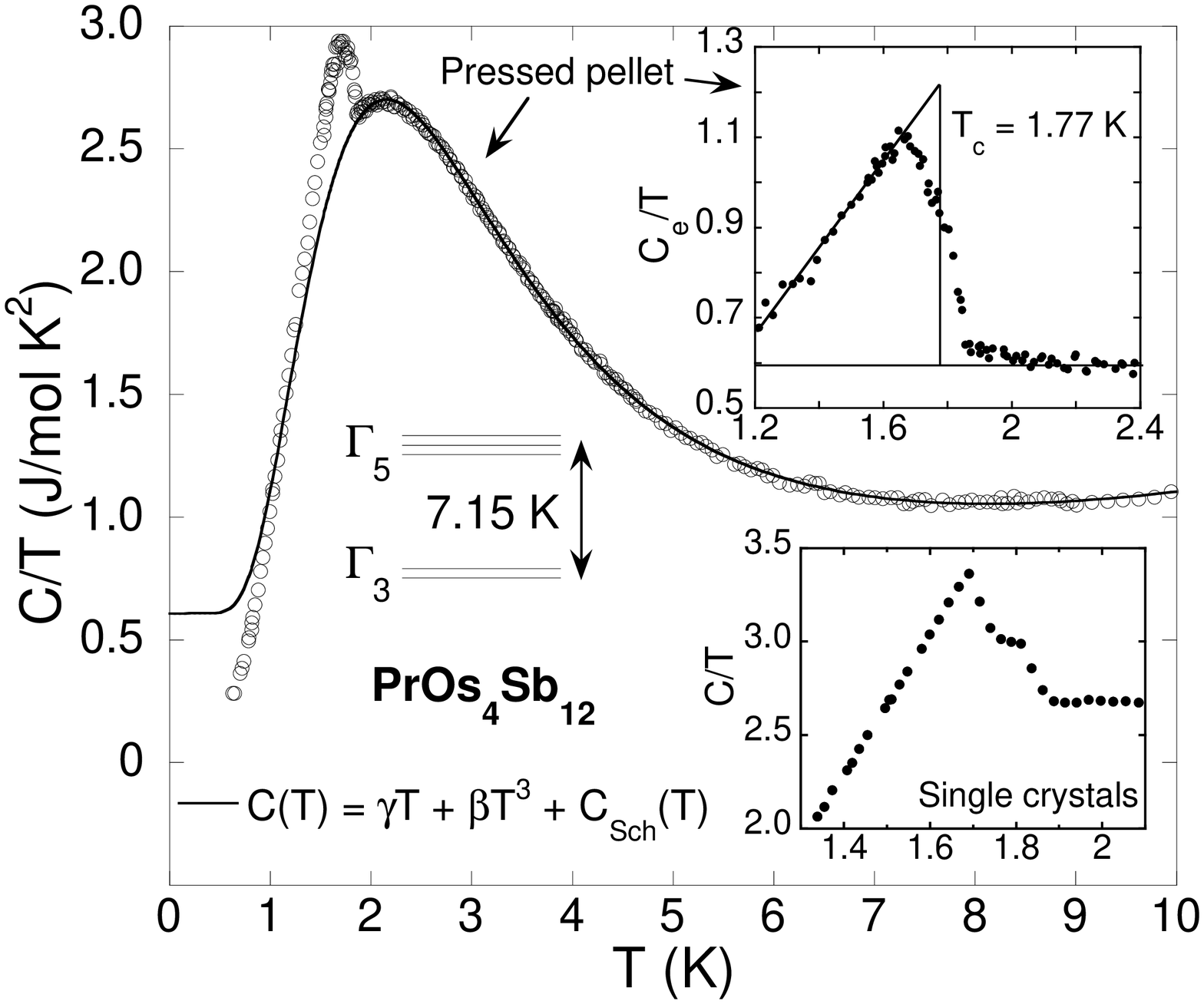} 
\caption{Specific heat $C$ divided by temperature $T$, $C/T$, vs.
$T$ for a \PrOsSb{} pressed pellet.  The line represents a fit of
the sum of electronic, lattice, and Schottky contributions to the
data.  Upper inset: $C_\mathrm{e}/T$ vs $T$ near $T_\mathrm{c}$
for a \PrOsSb{} pressed pellet ($C_\mathrm{e}$ is the electronic
contribution to $C$).  Lower inset: $C/T$ vs $T$ near
$T_\mathrm{c}$ for \PrOsSb{} single crystals, showing the
structure in $\Delta C$ near $T_\mathrm{c}$. Data from Ref.
\cite{Maple01,Bauer02}.} \label{heat}
\end{center}
\end{figure}

The first evidence that the superconductivity of the filled
skutterudite compound \PrOsSb{} develops out of a heavy Fermi
liquid emerged from measurements of the temperature dependence of
the specific heat $C(T)$.  Specific heat data from Refs.
\cite{Maple01} and \cite{Bauer02} in the form of a plot of $C/T$
vs $T$ between $0.5$ K and $10$ K for a \PrOsSb{} pressed pellet
(formed by pressing a collection of small single crystals in a
cylindrical die) are shown in Fig.~\ref{heat}.  The $C(T)$ data
have been corrected for Sb inclusions derived from the molten Sb
flux in which the crystals were grown.  The line in the figure
represents the expression $C(T) = \gamma T + \beta T^{3} +
C_\mathrm{Sch}(T)$, where $\gamma T$ and $\beta T^{3}$ are
electronic and phonon contributions, respectively, and
$C_\mathrm{Sch}(T)$ represents a Schottky anomaly for a two level
system consisting of a doublet ground state and a triplet excited
state at an energy $\Delta$ above the ground state.  The best fit
of this expression to the data yields the values $\gamma$ = $607$
mJ/mol K$^{2}$, $\beta = 3.95$ mJ/mol K$^{4}$ (corresponding to a
Debye temperature $\theta_\mathrm{D} = 203$ K), and $\Delta =
7.15$ K. Superimposed on the Schottky anomaly is a feature in the
specific heat due to the onset of superconductivity at
$T_\mathrm{c} = 1.85$ K which is also observed as an abrupt drop
in $\rho(T)$ to zero and as a sharp onset of diamagnetism in
$\chi(T)$. The feature in $C(T)/T$ due to the superconductivity is
also shown in the top inset of Fig.~\ref{heat} along with an
entropy conserving construction from which the ratio of the jump
in specific heat $\Delta C$ at $T_\mathrm{c}$ to $T_\mathrm{c}$,
$\Delta C/T_\mathrm{c} = 632$ mJ/mol K$^{2}$, has been estimated.
Using the BCS relation $\Delta C/k_\mathrm{B} T_\mathrm{c} =
1.43$, this yields an estimate for $\gamma$ of $440$ mJ/mol
K$^{2}$.  This value is comparable to that inferred for the fit of
the $C/T$ vs $T$ data in the normal state above $T_\mathrm{c}$,
and indicative of heavy fermion behavior.  A similar analysis of
the $C(T)$ data taken at the University of Karlsruhe on several
single crystals of \PrOsSb{} prepared in our laboratory yielded
$\gamma = 313$ mJ/mol K$^{2}$, $\theta_\mathrm{D} = 165$ K,
$\Delta = 7 $ K, and $\Delta C/\gamma{}T_\mathrm{c} \approx 3$,
much higher than the BCS value of 1.43 and indicative of strong
coupling effects \cite{Vollmer02}. Although the values of $\gamma$
determined from these experiments vary somewhat, they are all
indicative of a heavy electron ground state and an effective mass
$m^{*} \approx 50~m_\mathrm{e}$.

\begin{figure}[b]
\begin{center}\leavevmode
\includegraphics[width=0.8\linewidth]{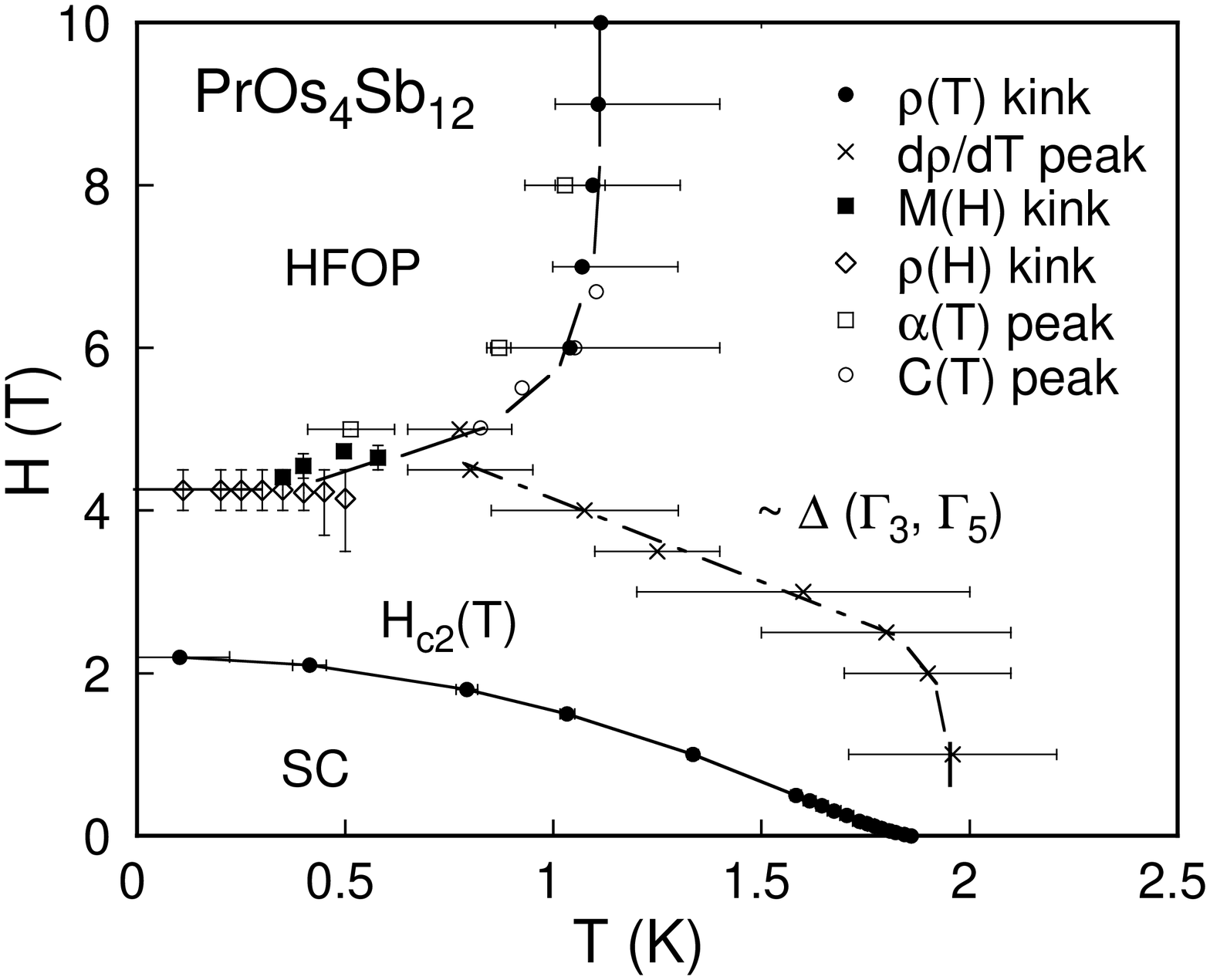} 
\caption{Magnetic field - temperature ($H - T$) phase diagram of
\PrOsSb{} showing the regions containing the superconductivity
(SC) and the high field ordered phase (HFOP). The dashed line is a
measure of the splitting between the Pr$^{3+} \Gamma_{3}$ ground
state and $\Gamma_{5}$ excited state (see text for further
details). Data from Refs.
\cite{Maple02,Ho01,Vollmer02,Oeschler02,Ho02}.} \label{phase}
\end{center}
\end{figure}

Further evidence of heavy fermion superconductivity is provided by
the upper critical field $H_\mathrm{c2}$ vs $T$ curve shown in
Fig.~\ref{phase} \cite{Bauer02,Maple02}.  The slope of the
$H_\mathrm{c2}$ curve near $T_\mathrm{c}$ can be used to determine
the orbital critical field $H^{*}_\mathrm{c2}(0) = \Phi_{0}/2\pi
\xi_{0}^{2}$, where $\Phi_{0}$ is the flux quantum, and, in turn,
the superconducting coherence length $\xi_{0}$, yielding the value
$\xi_{0} \approx 116$ \AA. The BCS relation $\xi_{0} = 0.18 \hbar
v_\mathrm{F}/k_\mathrm{B}T_\mathrm{c}$ can be used to estimate the
Fermi velocity $v_\mathrm{F}$ and, in turn, the effective mass
$m^{*}$ by means of the relation $m^{*} = \hbar
k_\mathrm{F}/v_\mathrm{F}$. Using a simple free electron model to
estimate the Fermi wave vector $k_{F}$, an effective mass $m^{*}
\approx 50~m_\mathrm{e}$ has been obtained \cite{Bauer02,Maple02}.
Calculating $\gamma$ from $m^{*}$ yields $\gamma \sim 350$ mJ/mol
K$^{2}$, providing further evidence for heavy fermion
superconductivity in \PrOsSb.

\section{Nature of the superconducting state}

Several features in the superconducting properties of \PrOsSb{}
indicate that the superconductivity of this compound is
unconventional in nature.  First, $C(T)$ follows a power law
T-dependence, $C_\mathrm{s}(T) \sim T^{2.5}$, after the Schottky
anomaly and $\beta T^{3}$ lattice contributions have been
subtracted from the $C(T)$ data.  Second, there is a `double-step'
structure in the jump in $C(T)$ near $T_\mathrm{c}$ in single
crystals (lower inset of Fig.~\ref{heat}) that suggests that there
are two distinct superconducting phases with different
$T_\mathrm{c}$'s: $T_\mathrm{c1} \approx 1.70$ K and
$T_\mathrm{c2} \approx 1.85$ K \cite{Maple02,Vollmer02}.  This
structure is not evident in the $C(T)$ data near $T_\mathrm{c}$
taken on the pressed pellet of \PrOsSb{} shown in the upper inset
of Fig.~\ref{heat}, possibly due to strains in the single crystals
out of which the pressed pellet is comprised that broaden the
transitions at $T_\mathrm{c1}$ and $T_\mathrm{c2}$ so that they
overlap and become indistinguishable. However, at this point, it
is not clear whether these two apparent jumps in $C(T)$ are
associated with two distinct superconducting phases or are due to
sample inhomogeneity.  It is noteworthy that all of the single
crystal specimens prepared in our laboratory and investigated by
our group and our collaborators exhibit this `double-step'
structure. Multiple superconducting transitions, apparently
associated with distinct superconducting phases, have previously
been observed in two other heavy fermion superconductors,
UPt$_{3}$ and U$_{1-x}$Th$_{x}$Be$_{13}~(0.1 \leq x \leq 0.35)$.
Measurements of the specific heat in magnetic fields reveal that
the two superconducting features shift downward in temperature at
nearly the same rate with increasing field, consistent with the
smooth temperature dependence of the $H_\mathrm{c2}(T)$ curve.
These two transitions have also been observed in thermal expansion
measurements \cite{Oeschler02}, which, from the Ehrenfest
relation, reveal that $T_\mathrm{c1}$ and $T_\mathrm{c2}$ have
considerably different pressure dependencies, suggesting that they
are associated with two distinct superconducting phases.

Recent transverse field $\mu$SR \cite{MacLaughlin02} and Sb-NQR
measurements \cite{Kotegawa02} on \PrOsSb{} are consistent with an
isotropic energy gap. Along with the specific heat, these
measurements indicate strong coupling superconductivity. These
findings suggest an s-wave, or, perhaps, a Balian-Werthamer p-wave
order parameter. Recently, the superconducting gap structure of
\PrOsSb{} was investigated by means of thermal conductivity
measurements in magnetic fields rotated relative to the
crystallographic axes by Izawa et al. \cite{Izawa02}.  These
measurements reveal two regions in the $H-T$ plane, a low field
region in which $\Delta(\bf{k})$ has two point nodes, and a high
field region where $\Delta(\bf{k})$ has six point nodes.  The line
lying between the low and high field superconducting phases may be
associated with the transition at $T_\mathrm{c2}$, whereas the
line between the high field phase and the normal phase,
$H_\mathrm{c2}(T)$, converges with $T_\mathrm{c1}$ as H
$\rightarrow 0$.  Clearly, more research will be required to
further elucidate the nature of the superconducting state in
\PrOsSb.

\section{Nature of the nonmagnetic heavy fermion state}

Magnetic susceptibility $\chi (T)$ measurements on \PrOsSb{}
indicate that it has a nonmagnetic ground state.  According to
Lea, Leask, and Wolf \cite{Lea62}, in a cubic CEF, the Pr$^{3+} J
= 4$ Hund's rule multiplet splits into a $\Gamma_{1}$ singlet, a
$\Gamma_{3}$ nonmagnetic doublet that carries an electric
quadrupole moment, and $\Gamma_{4}$ and $\Gamma_{5}$ triplets.  In
order to analyze the $\chi (T)$ data, it was assumed that the
Pr$^{3+}$ ions are, to first approximation, noninteracting and the
nonmagnetic ground state corresponds to either a $\Gamma_{1}$
singlet or $\Gamma_{3}$ nonmagnetic doublet ground state
\cite{Bauer02}.  Although reasonable fits to the $\chi (T)$ data
could be obtained for both $\Gamma_{1}$ and $\Gamma_{3}$ ground
states, the most satisfactory fit was obtained for a $\Gamma_{3}$
nonmagnetic doublet ground state with a $\Gamma_{5}$ first excited
triplet state at $11$ K and $\Gamma_{4}$ and $\Gamma_{1}$ second
and third excited states at 130 K and 313 K, respectively.
Inelastic neutron scattering measurements on \PrOsSb{} reveal
transitions at $0.71$ meV ($8.2$ K) and $11.5$ meV ($133$ K) that
appear to be associated with transitions between the $\Gamma_{3}$
ground state and the $\Gamma_{5}$ first and $\Gamma_{4}$ second
excited states, respectively, that are in good agreement with the
Pr$^{3+}$ CEF energy level scheme determined from the analysis of
the $\chi{}(T)$ data.  As noted above, the Schottky anomaly in the
$C(T)$ data on \PrOsSb{} taken at UCSD and at the University of
Karlsruhe \cite{Vollmer02} can be described well by a two level
system consisting of a doublet ground state and a low lying
triplet excited state with a splitting of $\sim 7$ K, a value that
is comparable to the values deduced from the $\chi (T)$ and INS
data. However, a $\Gamma_{1}$ ground state cannot, at this point,
be completely excluded.

While a magnetic $\Gamma_{4}$ or $\Gamma_{5}$ Pr$^{3+}$ ground
state could produce a nonmagnetic heavy fermion ground state via
an antiferromagnetic exchange interaction (Kondo effect), the
behavior of $\rho{}(T)$ of \PrOsSb{} in the normal state does not
resemble the behavior of $\rho{}(T)$ expected for this scenario.
For a typical magnetically induced heavy fermion compound,
$\rho{}(T)$ often increases with decreasing temperature due to
Kondo scattering, reaches a maximum, and then decreases rapidly
with decreasing temperature as the highly correlated heavy fermion
state forms below the coherence temperature.  At low temperatures,
$\rho{}(T)$ varies as $AT^{2}$ with a prefactor $A \approx
10^{-5}$ [$\mu{}\Omega$~cm~K$^{2}$(mJ/mol)$^{-2}$] $\gamma^{2}$
that is consistent with the Kadowaki-Woods relation
\cite{Kadowaki86}. In contrast, \PrOsSb{} has a very typical
metallic resistivity with negative curvature at higher
temperatures and a pronounced `roll off' below $\sim 8$ K before
it vanishes abruptly when the compound becomes superconducting.
The `roll off' in $\rho{}(T)$ appears to be due to a decrease in
charge or spin dependent scattering from the low lying Pr$^{3+}$
CEF energy level due to the decrease in population of this level
as the temperature is lowered.  The $\rho{}(T)$ data can be
described by a temperature dependence of the form $AT^{2}$ between
$\sim 8$ K and $45$ K, but with a prefactor $A \approx
0.009~\mu{}\Omega$~cm/K$^{2}$ that is nearly two orders of
magnitude smaller than that expected from the Kadowaki-Woods
relation which yields $A \approx 1.2~\mu\Omega$~cm/K$^{2}$ for
$\gamma \approx 350$ mJ/mol K$^{2}$ \cite{Kadowaki86}.
(Interestingly, $\rho{}(T)$ is consistent with $T^{2}$ behavior
with a value $A \approx 1~\mu\Omega$~cm/K$^{2}$ in fields of $\sim
5$ T in the high field ordered phase discussed in section $5$.)
However, the temperature dependence of $\rho (T)$ is similar to
that observed for the compound PrInAg$_{2}$ which has an enormous
$\gamma$ of $\sim 6.5$ J/mol K$^{2}$ and a $\Gamma_{3}$
nonmagnetic doublet ground state \cite{Yatskar96}. The compounds
\PrOsSb, PrInAg$_{2}$, and another Pr-based skutterudite,
PrFe$_{4}$P$_{12}$ \cite{Sato00}, may belong to a new class of
heavy fermion compounds in which the heavy fermion state is
produced by electric quadrupole fluctuations.  In contrast,
magnetic dipole fluctuations are widely believed to be responsible
for the heavy fermion state in most Ce and U heavy fermion
compounds (with the possible exception of certain U compounds such
as UBe$_{13}$). Another possible source of the enhanced effective
mass in \PrOsSb{} may involve excitations from the ground state to
the the low lying first excited state in the Pr$^{3+}$ CEF energy
level scheme \cite{Fulde02}.

\section{High field ordered phase}

Evidence for a high field ordered phase was first derived from
magnetoresistence measurements in the temperature range $80$ mK
$\leq T \leq 2$ K and magnetic fields up to $9$ tesla
\cite{Maple02,Ho02}. The $H-T$ phase diagram, showing the
superconducting region and the high field ordered phase is shown
in Fig.  2.  The line that intersects the high field ordered state
represents the inflection point of the `roll-off' in $\rho(T)$ at
low temperatures and is a measure of the splitting between the
Pr$^{3+}$ ground and the first excited states which decreases with
field.  The high field ordered phase has also been observed by
means of large peaks in the specific heat \cite{Vollmer02,Aoki02}
and thermal expansion \cite{Oeschler02} and kinks in magnetization
vs magnetic field curves \cite{Ho02,Tenya02} in magnetic fields $>
4.5$ T and temperatures $< 1.5$ K.

\section{Summary}

Experiments on the filled skutterudite compound \PrOsSb{} have
revealed a number of extraordinary phenomena: a heavy fermion
state characterized by an effective mass $m^{*} \approx
50~m_\mathrm{e}$, unconventional superconductivity below
$T_\mathrm{c}$ = $1.85$ with two distinct superconducting phases,
and a high field phase, presumably associated with magnetic or
electric quadrupolar order. Analysis of $\chi{}(T)$, $C(T)$,
$\rho{}(T)$, and INS data indicate that Pr$^{3+}$ has a
nonmagnetic $\Gamma_{3}$ doublet ground state that carries an
electric quadrupole moment, a low lying $\Gamma_{5}$ triplet
excited state at $\sim 10$ K, and $\Gamma_{4}$ triplet and
$\Gamma_{1}$ singlet excited states at much higher energies.  This
suggests that the interaction between the quadrupole moments of
the Pr$^{3+}$ ions and the charges of the conduction electrons and
excitations between the $\Gamma_{3}$ ground state and $\Gamma_{5}$
low lying excited state may play an important role in generating
the heavy fermion state and superconductivity in this compound.
The heavy fermion state and unconventional superconductivity will
constitute a significant challenge for theoretical description
\cite{Miyake02}.

\section{Acknowledgements}

This research was supported by US DOE Grant No. DE-FG03-86ER-45230,
US NSF Grant No. DMR-00-72125, and the NEDO International Joint
Research Program.

%
%


%
%

%
%

\end{document}